\documentclass[11pt]{article}
\usepackage[T1]{fontenc}
\usepackage[latin9]{inputenc}
\usepackage[a4paper]{geometry}
\usepackage[active]{srcltx}
\usepackage{textcomp}
\usepackage{amsmath}
\usepackage{amssymb}
\usepackage{esint}

\makeatletter

\pdfoutput=1

\usepackage{textcomp}

\pdfoutput=1 

\usepackage{jheppub}

\usepackage{etoolbox}
    
    \patchcmd{\maketitle}{\@fpheader}{}{}{}

\usepackage{amsfonts}

\usepackage{tikz}

\renewcommand{\d}{\partial}

\renewcommand{\tilde}{\widetilde}

\def\cD{\mathcal{D}}

\def\cH{\mathcal{H}}

\def\cL{\mathcal{L}}
\def\cM{\mathcal{M}}

\title{\boldmath Revisiting the asymptotic dynamics of General Relativity on AdS$_3$}

\author[a]{Hern\'an A. Gonz\'alez,}
\author[b]{Javier Matulich,}
\author[c]{Miguel Pino,}
\author[d]{Ricardo Troncoso}

\affiliation[a]{Institute for Theoretical Physics, TU Wien, Wiedner Hauptstr.~8-10/136, A-1040 Vienna, Austria}
\affiliation[b]{Universit\'e Libre de Bruxelles and International Solvay Institutes,
ULB-Campus Plaine CP231, 1050 Brussels, Belgium}
\affiliation[c]{Departamento de F\'{i}sica, Universidad de Santiago de Chile, Avenida Ecuador 3493, Estaci\'on Central, 9170124, Santiago, Chile}
\affiliation[d]{Centro de Estudios Cient\'{i}ficos (CECs), Av. Arturo Prat 514, Valdivia, Chile}

\emailAdd{hgonzale@hep.itp.tuwien.ac.at}
\emailAdd{jmatulic@ulb.ac.be}
\emailAdd{miguel.pino.r@usach.cl}
\emailAdd{troncoso@cecs.cl}

\preprint{CECS-PHY-18/03}

\abstract{The dual dynamics of Einstein gravity on AdS$_3$ supplemented with boundary conditions of KdV-type 
is identified. It corresponds to a two-dimensional field theory at the boundary, described by a novel action principle 
whose field equations are given by two copies of the ``potential modified KdV'' equation. The asymptotic symmetries
 then transmute into the global Noether symmetries of the dual action, giving rise to an infinite set of commuting conserved charges, 
implying the integrability of the system. Noteworthy, the theory at the boundary is non-relativistic and possesses anisotropic scaling
of Lifshitz type.}

\makeatother

\begin{document}
\maketitle \flushbottom

\newpage{}

\section{Introduction}

The dynamics of Einstein gravity in three spacetime dimensions  is described by global degrees of freedom that can be  identified only once a precise set of boundary conditions is provided.  In the case of asymptotically AdS spacetimes equipped with Brown-Henneaux boundary
conditions, the asymptotic symmetry group is generated by two copies of the Virasoro
algebra \cite{Brown:1986nw}. Demanding that the Lagrange multipliers
\textendash given by the lapse and shift functions in an ADM foliation\textendash{}
are held constant at infinity, the reduced phase space of the Einstein
field equations is described by Virasoro modes $\cL^{\pm}$ that evolve
according to 
\begin{equation}
\pm\ell\d_{t}\cL^{\pm}=\d_{\phi}\cL^{\pm},\label{I1}
\end{equation}
where $\ell$ is the AdS radius and $t,\phi$ are coordinates parametrizing
the cylinder at infinity. The symmetry algebra and the form of the
latter equation is consistent with the description in terms of the
boundary theory; it is well\textendash known that the asymptotic dynamics
for these boundary conditions is described by left and right chiral
bosons \cite{Coussaert:1995zp,Henneaux:1999ib}\footnote{As shown in \cite{Coussaert:1995zp,Henneaux:1999ib}, it is possible to rewrite the
action of two chiral bosons as a Liouville theory. This is accomplished
by performing a Bäcklund transformation that excludes the zero mode
sector of the chiral bosons.}. The components of the stress\textendash energy tensor of the chiral
bosons are given by the Virasoro modes $\cL^{\pm}$, so that equation
\eqref{I1} corresponds to its conservation law. Note that for the
boundary conditions of Brown and Henneaux, the chiral bosons and their
corresponding left/right energies fulfill the same equations.

Recently, a new family of boundary conditions connecting Einstein
gravity on AdS$_{3}$ with the Korteweg-de Vries (KdV) hierarchy of
integrable systems has been proposed in \cite{Perez:2016vqo}. The
possible choices of boundary conditions are labeled by a nonnegative
integer $n$, corresponding to the $n$-th representative of the hierarchy.
The Brown-Henneaux boundary conditions are recovered for $n=0$, so
that the modes fulfill \eqref{I1}; while for $n=1$, the modes are
described by noninteracting movers, satisfying the KdV equation 
\begin{equation}
\pm\ell\d_{t}\cL^{\pm}=3\cL^{\pm}\d_{\phi}\cL^{\pm}-\frac{\ell}{16\pi G}\d_{\phi}^{3}\cL^{\pm},\label{I2}
\end{equation}
where $G$ is the Newton constant. For $n\geq1$ the asymptotic symmetry
algebra turns out to be spanned precisely by the infinite set of commuting
charges of KdV. 

One of the main purposes of our work, is to unveil the precise form
of the action principle that describes the dynamics of the underlying
fields of the dual theory at the boundary, from which the field equations
of the KdV hierarchy emerge from a conservation law. In order to carry
out this task, it is convenient to use the Chern-Simons formulation
of three-dimensional gravity \cite{Achucarro:1987vz,Witten:1988hc}.
We then perform a Hamiltonian reduction similar to the one of Coussaert,
Henneaux and van Driel \cite{Coussaert:1995zp}. A distinguishing
feature of our derivation is that, as the boundary conditions for
$n>0$ actually precludes one from passing through the standard Hamiltonian
reduction of the Wess\textendash Zumino\textendash Witten (WZW) model
\cite{Forgacs:1989ac,Alekseev:1988ce},
one has to circumvent this step through imposing the boundary conditions
in the action principle from scratch. In this way, one obtains a novel
action principle for the dual theory, whose field equations are described
by two copies of the hierarchy of ``potential modified KdV'' (pmKdV)
equations of opposite chirality\footnote{A list of the first four equations of the pmKdV  hierarchy is given
in appendix \ref{modi}.}.

The paper is organized as follows. In the next section, we revisit
the boundary conditions of KdV-type in the context of 3D gravity with
negative cosmological constant. In section \ref{dual}, the dual theory at the boundary is obtained from the Hamiltonian reduction of the Chern-Simons action endowed with a suitable boundary term. The field equations are also analyzed. Section \ref{symmetries} is devoted to study the global symmetries symmetries
of the dual action principle at the boundary. We conclude in  with some comments in section \ref{concl}.

\section{General Relativity on AdS$_{3}$ and the KdV hierarchy}

\label{i}

Three-dimensional gravity with negative cosmological constant can
be formulated as the difference of two Chern-Simons actions for $sl(2,\mathbb{R})$-valued
gauge fields $A^{\pm}$ \cite{Achucarro:1987vz,Witten:1988hc} 
\begin{equation}
I=I_{CS}[A^{+}]-I_{CS}[A^{-}]\,,\label{action}
\end{equation}
where $I_{CS}$ reads 
\begin{equation}
I_{CS}[A]=\frac{k}{4\pi}\int_{\cM}\langle AdA+\frac{2}{3}A^{3}\rangle\,,\label{cs}
\end{equation}
and the Chern-Simons level is given by $k=\frac{\ell}{4G}$. Here,
$\cM$ is the three-dimensional manifold with coordinates $t,r,\phi$,
where $t$ represents time, $r$ stands for the radial coordinate
and $\phi$ is an angle. The generators of the $sl(2,\mathbb{R})$
algebra, given by $L_{m}^{\pm}$, with $m=\{-1,0,1\}$, are chosen
such that the commutators and the invariant non-degenerate bilinear
form read 
\begin{equation}
[L_{m}^{\pm},L_{n}^{\pm}]=(m-n)L_{m+n}^{\pm},\quad[L_{m}^{\pm},L_{n}^{\mp}]=0,
\end{equation}
and 
\begin{equation}
\langle{L_{0}^{\pm}}^{2}\rangle=\frac{1}{2},\quad\langle L_{1}^{\pm}L_{-1}^{\pm}\rangle=-1,
\end{equation}
respectively.

In order to describe the asymptotic form of the gauge fields, it is
useful to make a gauge choice as in \cite{Coussaert:1995zp}, so that
the connection reads
\begin{equation}
A^{\pm}=b_{\pm}^{-1}a^{\pm}b_{\pm}+b_{\pm}^{-1}db_{\pm},\label{radialgauge}
\end{equation}
with $b_{\pm}=e^{\pm\log(r/l)L_{0}^{\pm}}$. The components of the
auxiliary connection $a^{\pm}=a_{\phi}^{\pm}d\phi+a_{t}^{\pm}dt$,
then depend only on time and the angular coordinate, and are generically
given by \cite{Henneaux:2013dra,Bunster:2014mua} 
\begin{equation}
\begin{split}a_{\phi}^{\pm} & =L_{\pm1}^{\pm}-\frac{2\pi}{k}\cL^{\pm}L_{\mp1}^{\pm},\\
a_{t}^{\pm} & =\pm\frac{1}{\ell}\mu^{\pm}L_{\pm1}^{\pm}-\frac{1}{\ell}\d_{\phi}\mu^{\pm}L_{0}^{\pm}\pm\left(\frac{1}{2\ell}\d_{\phi}^{2}\mu^{\pm}-\frac{2\pi}{k\ell}\cL^{\pm}\mu^{\pm}\right)L_{\mp1}^{\pm}\,,
\end{split}
\label{eq:kdvbc}
\end{equation}
where $\cL^{\pm}(t,\phi)$ stand for the dynamical fields, and $\mu^{\pm}(t,\phi)$
correspond to the Lagrange multipliers. In the asymptotic region,
the field equations, $F^{\pm}=dA^{\pm}+A^{\pm}\wedge A^{\pm}=0$,
reduce to 
\begin{equation}
\pm\ell\d_{t}\cL^{\pm}=\cD^{\pm}\mu^{\pm}\,,\label{eomD}
\end{equation}
where the operators $\cD^{\pm}$ are defined by
\begin{equation}
\cD^{\pm}\equiv(\d_{\phi}\cL^{\pm})+2\cL^{\pm}\d_{\phi}-\frac{k}{4\pi}\d_{\phi}^{3}\,.
\end{equation}

The asymptotic symmetries can then be explicitly found by demanding
the preservation of the auxiliary connection $a^{\pm}$ under gauge
transformations, $\delta a^{\pm}=d\eta^{\pm}+[a^{\pm},\eta^{\pm}]$,
where $\eta^{\pm}$ is a Lie-algebra-valued parameter. Thus, the asymptotic
form of $a_{\phi}$ is maintained for gauge transformations spanned
by parameters of the form 
\begin{equation}
\eta^{\pm}=\varepsilon^{\pm}L_{\pm1}^{\pm}\mp\d_{\phi}\varepsilon^{\pm}L_{0}^{\pm}+\left(\frac{1}{2}\d_{\phi}^{2}\varepsilon^{\pm}-\frac{2\pi}{k}\cL^{\pm}\varepsilon^{\pm}\right)L_{\mp1}^{\pm}\,,\label{parameter}
\end{equation}
where $\varepsilon^{\pm}$ are arbitrary functions of $t$ and $\phi$,
provided that the dynamical fields $\cL^{\pm}$ transform as 
\begin{equation}
\delta\cL^{\pm}=\cD^{\pm}\varepsilon^{\pm}\,.\label{deltaL}
\end{equation}
Preserving the temporal component of the gauge field $a_{t}^{\pm}$
then implies the following condition for the variation of the Lagrange
multipliers

\begin{equation}
\delta\mu^{\pm}=\pm\ell\d_{t}\varepsilon^{\pm}+\d_{\phi}\mu^{\pm}\varepsilon^{\pm}-\mu^{\pm}\d_{\phi}\varepsilon^{\pm}.\label{deltaeta}
\end{equation}

It is worth stressing that the boundary conditions turn out to be
fully determined only once the precise form of the Lagrange multipliers
at the boundary is specified. The results of Brown and Henneaux \cite{Brown:1986nw}
are then recovered when the Lagrange multipliers are held constants
at infinity $\mu^{\pm}=1$. A simple generalization is obtained by choosing arbitrary functions of the coordinates, so that $\mu^{\pm}=\mu^{\pm}(t,\phi)$
are kept fixed at the boundary $(\delta\mu^{\pm}=0)$ \cite{Henneaux:2013dra,Bunster:2014mua}.
Different choices of boundary conditions, in which the Lagrange multipliers
are allowed to depend on the dynamical fields and their spatial derivatives,
were proposed in \cite{Perez:2016vqo}. Hereafter, we focus in a special
family of boundary conditions of KdV-type, being labeled by a non
negative integer $n$. In this scenario, the Lagrange multipliers
are chosen to be given by the $n$-th Gelfand-Dikii polynomial \cite{Gelfand:1975rn}
evaluated on $\cL^{\pm}$, i.e., 

\begin{equation}
\mu^{\pm}\equiv\mu_{n}^{\pm}=R_{n}^{\pm}[\cL^{\pm}]\,.\label{muR}
\end{equation}
The polynomials can be constructed by means of the following recursion
relation\footnote{Note that the normalization of the Gelfand-Dikii polynomials used
here differs from the one in \cite{Perez:2016vqo}. } 
\begin{equation}
\cD^{\pm}R_{n}^{\pm}=\d_{\phi}R_{n+1}^{\pm}\,.\label{recurrence}
\end{equation}
Thus, in the case of $n=0$ one obtains that $\mu_{0}^{\pm}=R_{0}^{\pm}=1$,
which reduces to the boundary conditions of Brown and Henneaux \cite{Brown:1986nw}.
In this case, equation \eqref{deltaeta} implies that the parameters
$\varepsilon^{\pm}$ are chiral, while the dynamical fields also do, since
the field equations \eqref{eomD} reduce to \eqref{I1}. The next
case corresponds to $n=1$$,$ so that the choice of Lagrange multipliers
is given by $\mu_{1}^{\pm}=R_{1}^{\pm}=\cL^{\pm}$, and hence, the
field equations in \eqref{eomD} reduce to KdV 
\begin{equation}
\pm\ell\d_{t}\cL^{\pm}=3\cL^{\pm}\d_{\phi}\cL^{\pm}-\frac{k}{4\pi}\d_{\phi}^{3}\cL^{\pm}\,.\label{KdVeq}
\end{equation}
In the remaining cases, $n>1$, the field equations are then given
by the ones of the $n$-th representative of the KdV hierarchy.


Note that for $n>0$, the Lagrange multipliers acquire a non-trivial
variation at infinity. Nonetheless, as shown in \cite{Perez:2016vqo}
and further explained in the next section, the action principle can
be well defined because each of the Gelfand-Dikii polynomials $R_{n}^{\pm}$
can be expressed in terms of the variation of a functional, i.e., 
\begin{equation}
R_{n}^{\pm}=\frac{\delta H_{n}^{\pm}}{\delta\cL^{\pm}}\,,\quad H_{n}^{\pm}=\int d\phi\,\cH_{n}^{\pm}\,,\label{h1}
\end{equation}
where $H_{n}^{\pm}$ stand for the conserved quantities of KdV, and
$\cH^{\pm}_{n}$ are the corresponding densities \footnote{A list with the first Gelfand-Dikii polynomials, conserved quantities
of KdV and the corresponding field equations of the KdV hierarchy is given
in appendix \ref{GDandHam}.}. Furthermore, equation \eqref{deltaeta} becomes a consistency relation
for the time derivative of the asymptotic symmetry parameters $\varepsilon^{\pm}$.
Thus, for $n\geq1$, assuming that the parameters  depend exclusively on the dynamical fields and their spatial derivatives, but not explicitly  on the coordinates, the general solution of the consistency relation is given by a linear combination of the form 
\begin{equation}
\varepsilon^{\pm}=\sum_{j=0}^{\infty}\epsilon^{\pm}_{j}R_{j}^{\pm},\label{sumRn}
\end{equation}
with $\epsilon^{\pm}_{j}$ constants. This infinite set of symmetries then
gives rise to conserved charges, which can be written as surface integrals
by means of the Regge-Teitelboim approach \cite{Regge:1974zd}. The
variation of the conserved charges associated to the gauge transformation
generated by a parameter of the form \eqref{parameter} that spans the asymptotic symmetries, is given
by 
\begin{align}
\delta Q[\varepsilon^{\pm}]=\mp\int d\phi\;\varepsilon^{\pm}\delta\cL^{\pm}\,,\label{varQ}
\end{align}
which can be integrated due to \eqref{sumRn} and \eqref{h1}, yielding
\begin{equation}
Q[\varepsilon^{\pm}]=\mp\sum_{j=0}^{\infty}\epsilon^{\pm}_{j}H_{j}^{\pm}\,.
\label{cargae}
\end{equation}

The asymptotic symmetries are then canonically realized. A straightforward
way to obtain the asymptotic symmetry algebra in terms of Poisson
brackets is given by the relation 
\begin{equation}
\{Q[\varepsilon_{1}],Q[\varepsilon_{2}]\}=\delta_{\varepsilon_{2}}Q[\varepsilon_{1}]\,.\label{BHrel}
\end{equation}
The cases $n=0$ and $n>0$ are then very different in this context.
Indeed, for $n>0$ the algebra turns out to be abelian 
\begin{equation}
\{H_{i}^{\pm},H_{j}^{\pm}\}=0\,,\label{hk}
\end{equation}
while for $n=0$, which corresponds to Brown-Henneaux, the algebra
of the conserved charges is given by two copies of the Virasoro algebra
with a non-vanishing central extension.

Some interesting remarks about the metric formulation are in order.
It is worth highlighting that the reduced phase space for the boundary
conditions of KdV-type, for an arbitrary non negative integer $n$,
always contain the BTZ black hole \cite{Banados:1992wn,Banados:1992gq},
which corresponds to the configuration with $\mathcal{L}^{\pm}$ constants
\cite{Perez:2016vqo}. Indeed, the field equations of the KdV hierarchy
are trivially solved in this case, and the spacetime metric in the
ADM decomposition is such that the lapse and the shift correspond
to a non-standard foliation, determined by $\mu_{n}^{\pm}=\frac{\left(2n\right)!}{2^{n}\left(n!\right)^{2}}\left(\mathcal{L}^{\pm}\right)^{n}$.
Specifically 
\begin{align}
ds^{2} & =\ell^{2}\left[\frac{dr^{2}}{r^{2}}+\frac{2\pi}{k}\mathcal{L}^{+}\left(d\tilde{x}^{+}\right)^{2}+\frac{2\pi}{k}\mathcal{L}^{-}\left(d\tilde{x}^{-}\right)^{2}-\left(\frac{r^{2}}{\ell^{2}}+\left(\frac{2\pi\ell}{k}\right)^{2}\frac{\mathcal{L}^{+}\mathcal{L}^{-}}{r^{2}}\right)d\tilde{x}^{+}d\tilde{x}^{-}\right]\;,
\end{align}
with 
\begin{equation}
d\tilde{x}^{\pm}=\frac{1}{\ell}\mu_{n}^{\pm}dt\pm d\varphi\,.
\end{equation}
Furthermore, the boundary conditions described by \eqref{radialgauge}
and \eqref{eq:kdvbc}, with $\mu_{n}^{\pm}$ given by \eqref{muR}
are such that the fall-off of the metric somewhat resembles the one
of Brown-Henneaux. Indeed, in a Fefferman-Graham-like gauge, the spatial
components of the metric and its conjugate momenta behave as 
\begin{align}
g_{rr}=\frac{\ell^{2}}{r^{2}}\,, & \quad g_{r\phi}=0\,,\quad g_{\phi\phi}=r^{2}+O(1),\\
\pi^{rr}=O(r^{-1})\,, & \quad\pi^{r\phi}=O(r^{-2}),\quad\pi^{\phi\phi}=0\,.
\end{align}
However, the key difference arises in the asymptotic behavior of the
lapse and shift functions, which read
\begin{equation}
\begin{split}N^{\perp} & =\frac{1}{2}(\mu_{n}^{+}+\mu_{n}^{-})\frac{r}{\ell}+O(r^{-1})\,,\\
N^{r} & =-(\d_{\phi}\mu_{n}^{+}-\d_{\phi}\mu_{n}^{-})\frac{r}{2\ell}+O(r^{-2})\,,\\
N^{\phi} & =\frac{1}{2\ell}(\mu_{n}^{+}-\mu_{n}^{-})+O(r^{-2})\,.
\end{split}
\end{equation}
Hence, for $n>0$ they are allowed to fluctuate at leading order,
in sharp contrast with the fall-off for $n=0$ that corresponds to
the Brown-Henneaux boundary conditions for which $\mu_{0}^{+}=\mu_{0}^{-}=1$.

\section{Dual theory at the boundary}
\label{dual}

In this section, we perform a Hamiltonian reduction of the action
\eqref{action} by explicitly solving the constraints of the theory.
The boundary conditions for the gauge field $A^{\pm}$ correspond
to \eqref{radialgauge} and \eqref{eq:kdvbc}, where the ``chemical
potentials'' $\mu^{\pm}$ in \eqref{muR} are given by the $n$-th
Gelfand-Dikii polynomial $R_{n}^{\pm}[\cL^{\pm}]$. The reduction
is carried out for a generic value of $n$.

\subsection{Hamiltonian reduction}

The Hamiltonian reduction of Chern-Simons theory in the context of
three-dimensional gravity has been discussed extensively in the literature,
see e.g., \cite{Coussaert:1995zp,Henneaux:1999ib,Rooman:2000zi,Barnich:2013yka}.
For the standard choices of boundary conditions \cite{Brown:1986nw,Barnich:2006av}, the
classical dynamics
can be obtained from the  Hamiltonian reduction of the WZW theory at the boundary  \cite{Witten:1988hf,Elitzur:1989nr,Forgacs:1989ac,Alekseev:1988ce}. Nonetheless, for the boundary conditions of KdV-type, the
reduction does not lead to the usual WZW theory at the boundary, since for a generic value
of $n$ the components of the gauge field at the boundary are no longer
proportional, and hence, the Kac-Moody symmetry appears to be manifestly
broken (except when $n=0$ which corresponds to Brown-Henneaux). Nevertheless,
as explained below, the reduction can still be successfully performed
because the boundary conditions can be appropriately implemented in
the action principle.

The resulting reduced action at the boundary gives rise to a different
hierarchy of integrable equations, labeled by the integer $n$. The
simplest case ($n=0$) corresponds to two chiral bosons of opposite
chirality \cite{Coussaert:1995zp,Henneaux:1999ib}. For $n=1$ we
obtain a novel action principle, whose field equations are given by
two copies of the 
pmKdV equation (see e.g. \cite{olver2000applications,wang2002list}). In the remaining cases ($n>1$) the action of the dual theory
describes the other members of the pmKdV hierarchy. The integrability
of this hierarchy is explicitly checked the next section. 

We start with the action \eqref{action} written in explicit Hamiltonian
form 
\begin{equation}
I=I_{H}[A^{+}]-I_{H}[A^{-}]\,,\label{actionH}
\end{equation}
with 
\begin{equation}
I_{H}[A^{\pm}]=\frac{k}{4\pi}\int dtd^{2}x\;\epsilon^{ij}\langle\dot{A}_{i}^{\pm}A_{j}^{\pm}+A_{t}^{\pm}F_{ij}^{\pm}\rangle+B^{\pm}\,,\label{IHpm}
\end{equation}
where $B^{\pm}$ stand for appropriate boundary terms generically
needed in order to have an action principle that is well defined.
It is worth pointing out that the boundary can be located at an arbitrary
fixed value of the radial coordinate. Here $\epsilon^{ij}$ is the
spatial part of the Levi-Civita symbol, while $F_{ij}^{\pm}$ is the
curvature $F_{ij}^{\pm}=\d_{i}A_{j}^{\pm}-\d_{j}A_{i}^{\pm}+[A_{i}^{\pm},A_{j}^{\pm}]$.
We choose $\epsilon^{r\phi}=1$, and dot stands for derivative with respect to $t$. The action \eqref{IHpm} attains an extremum when
the field equations hold, provided that 
\begin{equation}
\delta B^{\pm}=-\frac{k}{2\pi}\int d\phi dt\;\langle A_{t}^{\pm}\delta A_{\phi}^{\pm}\rangle\,.
\end{equation}
Note that for the Brown-Henneaux boundary conditions ($n=0$), the
components of the gauge field satisfy $\ell A_{t}^{\pm}=A_{\phi}^{\pm}$
at the boundary, and hence, $\delta B^{\pm}$ can be readily integrated.
However, for the boundary conditions of KdV-type, with $n\geq1\text{, }$the
temporal and angular components of the gauge field at the boundary
are not proportional (see \eqref{eq:kdvbc}), and so one might worry
about the integrability of the boundary terms $B^{\pm}$. However,
as explained in \cite{Perez:2016vqo}, since the Lagrange multipliers
$\mu^{\pm}$ in \eqref{muR} are given by the variation of a functional
(see \eqref{h1}) the boundary terms can be explicitly integrated
as 
\begin{equation}
B^{\pm}=\mp\frac{1}{\ell}\int d\phi dt\;\cH_{n}^{\pm}\,.
\end{equation}

Therefore, the suitable action principle for the boundary conditions
of KdV-type is precisely identified, and so we are able to proceed
with its Hamiltonian reduction. 

The constraint $\epsilon^{ij}F_{ij}^{\pm}=0$ is locally solved by
$A_{i}^{\pm}=G_{\pm}^{-1}\d_{i}G_{\pm}$. For the sake of simplicity,
we disregard non-trivial holonomies, so that $G_{\pm}(t,r,\phi)\in$
$SL(2,\mathbb{R})$ can be assumed to be periodic in $\phi$. Thus,
replacing back in the action \eqref{IHpm}, a straightforward calculation
yields

\begin{equation}
I_{H}[A^{\pm}]=I_{1}^{\pm}+I_{2}^{\pm}+B^{\pm}\,,\label{I_H}
\end{equation}
where 
\begin{align}
I_{1}^{\pm}= & \frac{k}{4\pi}\int dtdrd\phi\;\epsilon^{ij}\langle\d_{t}({G_{\pm}}^{-1})\d_{i}G_{\pm}G_{\pm}^{-1}\d_{j}G_{\pm}\rangle\,,\\
I_{2}^{\pm}= & -\frac{k}{4\pi}\int d\phi dt\;\langle\d_{t}G_{\pm}\d_{\phi}(G_{\pm}^{-1})\rangle.
\end{align}
The first two terms $I_{1}^{\pm}+I_{2}^{\pm}$ naturally appear in
the standard chiral WZW action \cite{Elitzur:1989nr}, but here we
have an explicit modification due to the presence of $B^{\pm}$. As shown below, the form of $B^{\pm}$ makes possible to recover the
infinite-dimensional Abelian algebra in \eqref{hk} from a Noether
symmetry of the full action. Furthermore, note that $B^{\pm}$ do
not appear to be expressible locally in terms of the group elements
$G_{\pm}$.


In order to reduce $I_{1}^{\pm}$ to a boundary integral, we use the
Gauss decomposition for $G_{\pm}$ 
\begin{equation}
G_{\pm}=\exp\left[X_{\pm}L_{\pm1}^{\pm}\right]\exp\left[\pm\Phi_{\pm}L_{0}^{\pm}\right]\exp\left[Y_{\pm}L_{\mp1}^{\pm}\right].\label{gauss_capital}
\end{equation}
Here $X_{\pm}$, $Y_{\pm}$ and $\Phi_{\pm}$ are functions of $t,r,\phi$.
Thus, $I_{1}^{\pm}$ can be expressed as 
\begin{equation}
I_{1}^{\pm}=\frac{k}{4\pi}\int d\phi dt\;e^{\Phi_{\pm}}(Y'_{\pm}\dot{X}_{\pm}-\dot{Y}_{\pm}X'_{\pm})\,,
\end{equation}
where prime denotes derivatives with respect to $\phi$. Thus, the
action $I_{H}[A^{\pm}]$ has now been reduced to an integral at the
boundary.

The fall-off in \eqref{radialgauge}, \eqref{eq:kdvbc} allows us
to decompose the group element $G_{\pm}$ in the asymptotic region
as a product of two group elements, according to $G_{\pm}(t,r,\phi)=g_{\pm}(t,\phi)b_{\pm}(r)$,
with $b_{\pm}=e^{\pm\log(r/\ell)L_{0}^{\pm}}$. Therefore, $I_{2}^{\pm}$
reduces to 
\begin{equation}
I_{2}^{\pm}=-\frac{k}{4\pi}\int d\phi dt\;\langle\dot{g}_{\pm}\d_{\phi}g_{\pm}^{-1}\rangle\,,
\end{equation}
which can be further simplified by performing the Gauss decomposition
for the group element $g_{\pm}$ 
\begin{equation}
g_{\pm}=e^{x_{\pm}L_{\pm1}^{\pm}}e^{\pm\varphi_{\pm}L_{0}^{\pm}}e^{y_{\pm}L_{\mp1}^{\pm}},\label{gauss}
\end{equation}
where the fields $x_{\pm}$, $y_{\pm}$ and $\varphi_{\pm}$ depend
only on $t$ and $\phi$. Thus, we obtain 
\begin{equation}
I_{2}^{\pm}=\frac{k}{4\pi}\int d\phi dt\;\left(\frac{1}{2}\dot{\varphi}_{\pm}\varphi_{\pm}'-e^{\varphi_{\pm}}(y'_{\pm}\dot{x}_{\pm}+\dot{y}_{\pm}x'_{\pm})\right).
\end{equation}
Furthermore, consistency of \eqref{gauss_capital} and \eqref{gauss}
yields
\begin{equation}
Y_{\pm}=\frac{\ell}{r}y_{\pm}\,,\quad X_{\pm}=x_{\pm}\,,\quad e^{\Phi_{\pm}}=\frac{r}{\ell}e^{\varphi_{\pm}}\,,\label{capital-non}
\end{equation}
and hence, the action \eqref{I_H} reduces to 
\begin{equation}
I_{H}[A^{\pm}]=\frac{k}{4\pi}\int d\phi dt\;\left(\frac{1}{2}\dot{\varphi}_{\pm}\varphi'_{\pm}-2e^{\varphi_{\pm}}x'_{\pm}\dot{y}_{\pm}\right)\mp\frac{1}{\ell}\int d\phi dt\;\cH_{n}^{\pm}\,.\label{Iaux1}
\end{equation}
Besides, the asymptotic form of $a_{\phi}^{\pm}=g_{\pm}^{-1}\d_{\phi}g_{\pm}$
is determined by eq. \eqref{eq:kdvbc}, so that
\begin{equation}
g_{\pm}^{-1}\d_{\phi}g_{\pm}=L_{\pm1}^{\pm}-\frac{2\pi}{k}\cL^{\pm}L_{\mp1}^{\pm}\,,
\end{equation}
which by virtue of the Gauss decomposition \eqref{gauss}, implies
the following relations 
\begin{eqnarray}
e^{\varphi_{\pm}}x'_{\pm}=1\,,\quad2e^{\varphi_{\pm}}y_{\pm}x'_{\pm}+\varphi'_{\pm}=0\,,\quad y'_{\pm}+e^{\varphi_{\pm}}y_{\pm}^{2}x'_{\pm}+y_{\pm}\varphi'_{\pm}=-\frac{2\pi}{k}\cL^{\pm}\,.\label{gd1}
\end{eqnarray}
Making use of the first equation in \eqref{gd1}, it is straightforward
to see that the second term in \eqref{Iaux1} becomes a total time
derivative that can be discarded. The remaining equations in \eqref{gd1}
then allow to obtain a crucial relationship, given by
\begin{equation}
\cL^{\pm}=\frac{k}{4\pi}\left(\frac{1}{2}{\varphi'_{\pm}}^{2}+\varphi''_{\pm}\right)\,,\label{miura}
\end{equation}
from which the reduced action \eqref{Iaux1} can be expressed exclusively
in term of two fundamental fields $\varphi_{\pm}$. 

In sum, the action of the dual theory at the boundary explicitly reads
\begin{equation}
\label{reduced0}
I_{n}[\varphi_+,\varphi_-]=I_{n}[\varphi_+]-I_{n}[\varphi_-]\,,
\end{equation}
with
\begin{equation}
I_{n}[\varphi_{\pm}]=\frac{k}{8\pi}\int d\phi dt\;\left(\dot{\varphi}_{\pm}\varphi'_{\pm}\mp\frac{8\pi}{\ell k}\cH_{n}^{\pm}[\cL^{\pm}]\right),\label{reduced}
\end{equation}
where $\cL^{\pm}$ is given by \eqref{miura}.  A complete analysis of the global symmetries of the dual action \eqref{reduced}
is performed in section \ref{symmetries}. 

For the remaining steps, it is worth highlighting that the action
\eqref{reduced} possesses the following gauge symmetry 
\begin{equation}
\varphi_{\pm}\rightarrow\varphi_{\pm}+f_{\pm}(t)\,,\label{gaugered}
\end{equation}
where $f_{\pm}$ stand for arbitrary functions. Indeed, under \eqref{gaugered},
the kinetic term in \eqref{reduced} just changes by a time derivative,
while the Hamiltonian does not give additional contributions since
$\cH_{n}^{\pm}$ only involves angular derivatives of $\varphi_{\pm}$.

In order to familiarize with the dynamics of the theory at the boundary, it is certainly useful to analyze its field equations. As
it is shown below, the gauge freedom in \eqref{gaugered} allows to
write down the equations of motion of \eqref{reduced} precisely as
those of the pmKdV hierarchy. 

\subsection{$n=0$: chiral bosons}

The simplest case is given by $\mu_{0}^{\pm}=1$, which corresponds
to the Brown-Henneaux boundary conditions. The dual theory is obtained
from \eqref{reduced0} and \eqref{reduced} with $n=0$, so that $\cH_{0}^{\pm}=\cL^{\pm}$,
and hence each copy of the action reads 
\begin{equation}
I_{0}[\varphi_{\pm}]=\frac{k}{8\pi}\int d\phi dt\;\left(\dot{\varphi}_{\pm}\varphi'_{\pm}\mp\frac{1}{\ell}{\varphi'_{\pm}}^{2}\right)\,,
\end{equation}
in agreement with the standard result obtained in \cite{Coussaert:1995zp}.
The theory describes the dynamics of two chiral bosons of opposite
chirality. The field equations in this case then read
\begin{equation}
\dot{\varphi}'_{\pm}=\pm\frac{1}{\ell}\varphi''_{\pm}\,,
\end{equation}
which can be readily integrated once, yielding 
\begin{equation}
\dot{\varphi}_{\pm}=\pm\frac{1}{\ell}\varphi'_{\pm}+h_{\pm}(t)\,,
\end{equation}
where $h_{\pm}$ are arbitrary functions of time. Therefore, these
arbitrary functions can be set to zero by virtue of the gauge symmetry
in \eqref{gaugered}, with $h_{\pm}(t)=\dot{f}_{\pm}(t)$, and hence
\begin{equation}
\dot{\varphi}_{\pm}=\pm\frac{1}{\ell}\varphi'_{\pm}\,.\label{eq:chiralboson}
\end{equation}
Note that, as mentioned in the introduction, the field equations for
$n=0$ in \eqref{eq:chiralboson} coincide with the ones of the Virasoro
modes in \eqref{I1}. As it is shown below, in our context, the fact
that the field equation is equivalent to the conservation law it is
actually an accident of the particular case $n=0$.

\subsection{$n=1$: pmKdV movers}

The next case corresponds to the choice $\mu_{1}^{\pm}=\cL^{\pm}$ so that 
$\cH_{1}^{\pm}=\frac{1}{2}{\cL^{\pm}}^{2}$. The chiral copies of the actions
then read
\begin{equation}
I_{1}[\varphi_{\pm}]=\frac{k}{8\pi}\int d\phi dt\;\left(\dot{\varphi}_{\pm}\varphi'_{\pm}\mp\frac{k}{16\pi\ell}{\varphi'_{\pm}}^{4}\mp\frac{k}{4\pi\ell}{\varphi''_{\pm}}^{2}\right)\,,
\end{equation}
and the field equations are given by
\begin{equation}
\dot{\varphi}'_{\pm}=\pm\frac{k}{4\pi\ell}\left(\frac{3}{2}{\varphi_{\pm}'}^{2}\varphi_{\pm}''-\varphi_{\pm}''''\right)\,.\label{mkdv}
\end{equation}
As in the previous case, the equation can be integrated once,
giving 
\begin{equation}
\dot{\varphi}_{\pm}=\pm\frac{k}{4\pi\ell}\left(\frac{1}{2}{\varphi_{\pm}'}^{3}-\varphi_{\pm}'''\right)\,,\label{pmkdv}
\end{equation}
where the arbitrary integration function has been set to zero by virtue of an
appropriate gauge choice. This equation corresponds to two copies
of the pmKdV equation.\footnote{The name stems from the fact that under the identification $u=\varphi'$,
equation \eqref{mkdv} reduces to modified KdV (mKdV) for $u$.}

It is worth highlighting that in this case the dynamics of the theory at the boundary, described by \eqref{pmkdv}, differs from the conservation
law that fulfills the KdV equation \eqref{I2}.

\subsection{Generic $n$: pmKdV hierarchy }
The generic choice of boundary conditions corresponds to $\mu^{\pm}_{n}=R^{\pm}_{n}=\frac{\delta H_{n}^{\pm}}{\delta\cL^{\pm}}$, where $R^{\pm}_{n}$ stand for the $n$--th Gelfand-Dikii  polynomials. The action is then given by \eqref{reduced0} and \eqref{reduced} with \eqref{miura}.

The field equations can be readily obtained in a closed form for a generic value of $n$, yielding
\begin{equation}
\dot{\varphi}'_{\pm}=\pm\frac{1}{\ell}\d_{\phi}\left(R_{n}^{\pm}\varphi'_{\pm}-\d_{\phi}R_{n}^{\pm}\right)\,.
\end{equation}
As in the previous cases, these equations can be integrated once, and by means of the gauge symmetry of the action \eqref{gaugered}, they reduce to
\begin{equation}
\dot{\varphi}_{\pm}=\pm\frac{1}{\ell}\left(R_{n}^{\pm}\varphi'_{\pm}-\d_{\phi}R_{n}^{\pm}\right)\,,\label{ksks}
\end{equation}
in agreement with $n$--th representative of the potential form of the mKdV
hierarchy.

In the next section it is shown that these equations 
can be manifestly seen to be integrable, since they admit an  infinite number of commuting conserved charges.

\section{Symmetries of the action}
\label{symmetries}

This section is devoted to study the symmetries and conserved
quantities of the action \eqref{reduced}. Apart from the gauge symmetry
\eqref{gaugered}, the action \eqref{reduced0} also possesses global and kinematic symmetries,
which are described in what  follows.

\subsection{Global symmetries}
Here we show that the action \eqref{reduced0} is invariant under the following Noether symmetries 
\begin{equation}
\delta\varphi_{\pm}=\varepsilon^{\pm}\varphi'_{\pm}-\varepsilon'^{\pm}\,,\label{transphi}
\end{equation}
with $\varepsilon^{\pm}$ given by  \eqref{sumRn}. It is worth stressing that these global symmetries are in one to one correspondence with the asymptotic symmetries in the bulk. Indeed, by means of the map in \eqref{miura}, the transformation law of $\cL^{\pm}$ that is given by \eqref{deltaL} is precisely recovered from \eqref{transphi}.
Therefore, the corresponding infinite number of commuting Noether charges $H_{n}^{\pm}$  can be seen to coincide with  the surface integrals that come from the analysis in the bulk.

This can be explicitly shown as follows. For each copy of the action \eqref{reduced0}, the Hamiltonian is invariant under transformation
\eqref{transphi}, while the kinetic term changes by a total derivative in time.
Indeed, equations \eqref{deltaL} and \eqref{sumRn}, imply that the variation of the Hamiltonian term can be expressed as
\begin{equation}
\delta\int d\phi dt\;\cH_{n}^{\pm}=\sum_{k=0}^{\infty}\epsilon^{\pm}_{k}\int d\phi dt\;\frac{\delta H_{n}^{\pm}}{\delta\cL^{\pm}}\cD^{\pm}\frac{\delta H_{k}^{\pm}}{\delta\cL^{\pm}}\,.
\end{equation}
Thus, the r.h.s. corresponds to the Poisson bracket $\{H^{\pm}_{n}, H^{\pm}_{k}\}$, which vanishes due to \eqref{recurrence}.
The variation of the kinetic term in \eqref{reduced0} reduces to
\begin{equation}
\delta\frac{k}{8\pi}\int d\phi dt\;\dot{\varphi}_{\pm}\varphi'_{\pm}=\int d\phi dt\left(-\sum_{k=0}^{\infty}\epsilon^{\pm}_{k}\dot{\cH}_{k}^{\pm}+\frac{k}{8\pi}\partial_{t}(\delta\varphi_{\pm}\varphi'_{\pm})\right)\,,
\end{equation}
and hence the transformation \eqref{transphi} is a symmetry of the action. Therefore, the straightforward
application of Noether's theorem yields an infinite number of commuting conserved charges given by 
\begin{equation}
Q(\varepsilon^{\pm})=\mp\sum_{k=0}^{\infty}\epsilon^{\pm}_{k}H_{k}^{\pm}\,,
\label{cargasn}
\end{equation}
which implies that the field equations in \eqref{ksks} correspond to an integrable system.
Besides, and noteworthy, the Noether charges associated with the global symmetries of the dual theory  in \eqref{cargasn} precisely agree with the surface integrals found from the asymptotic symmetries in the bulk \eqref{cargae}.

\subsection{Kinematic symmetries \& Lifshitz scaling }

The kinematic symmetries of the dual action \eqref{reduced0} correspond to rigid  displacements in space and time, as well as global anisotropic scaling. These symmetries are spanned by  a two-dimensional
vector field 
\begin{equation}
X=\left(\alpha^{t}+\gamma zt\right)\partial_{t}+\left(\alpha^{\phi}+\gamma\phi\right)\partial_{\phi}\,,\label{X}
\end{equation}
with $\alpha^{t}$, $\alpha^{\phi}$ and $\gamma$ constants, and $z$ is related
to the integer $n$ through $z=2n+1$. Under an infinitesimal
diffeomorphism spanned by $X$ the scalar fields transform as $\delta_{X}\varphi_{\pm}=X^{\mu}\d_{\mu}\varphi_{\pm}$,
which implies that left and right Hamiltonians change according to
\begin{equation}
\delta_{X}\cH_{n}^{\pm}[\cL^{\pm}]=X^{\mu}\d_{\mu}\cH_{n}^{\pm}+\gamma(z+1)\cH_{n}^{\pm}\,,\label{sca}
\end{equation}
so that under anisotropic scaling they have weight given by
$2n+2$.

It is simple to prove that the dual action \eqref{reduced} is invariant under the symmetries spanned by $X$, and hence, the corresponding Noether charges for the chiral copies are given by \begin{equation}
Q^{\pm}[\alpha^{t},\alpha^{\phi},\gamma]=\int d\phi\left[\left(\alpha^{\phi}+\gamma\phi\right)\cL^{\pm}\pm\frac{1}{\ell}\left(\alpha^{t}+\gamma zt\right)\cH_{n}^{\pm}\right]\,.
\end{equation}
Thus, for each copy, the energy, the momentum, and the conserved charge associated to anisotropic scaling are given by
\begin{equation}
H^{\pm}=Q^{\pm}[1,0,0]=\pm\frac{1}{\ell}H^{\pm}_n,\quad
P^{\pm}=Q^{\pm}[0,1,0]=H^{\pm}_0, \quad D^{\pm}=Q^{\pm}[0,0,1]\,,
\end{equation}
respectively. Note that $D^{\pm}$ correspond to left and right copies of the generator of anisotropic Lifshitz scaling of the form, 
\begin{equation}
t\mapsto\lambda^{z}t,\quad\phi\mapsto\lambda\phi\,,\label{symXl}
\end{equation}
so that $z=2n+1$ stands for the dynamical exponent. For each copy, the generators of the kinematic symmetry  then fulfill the Lifshitz algebra in two dimensions (see e.~g.
\cite{Hartnoll:2009sz,Gonzalez:2011nz,Taylor:2015glc}). In fact, as it can be readily obtain from relation \eqref{BHrel},
one obtains \footnote{In order to recover the Lifshitz algebra, it is useful to make use of the following identity \cite{McKean1975,0036-0279-31-1-R03,Gelfand:1975rn}: 
\[
(2n+1)H_{n}^{\pm}=\int d\phi\text{\,}R_{n+1}^{\pm}\:.
\]
} 
\begin{equation}
\{P^{\pm},H^{\pm}\}=0\,,\quad\{D^{\pm},P^{\pm}\}=P\,,\quad\{D^{\pm},H^{\pm}\}=zH^{\pm}\,.
\end{equation}
In summary, the only non-vanishing commutators of the infinite set of  global symmetries are given by 
\begin{equation}
\{D^{\pm},H_{j}^{\pm}\}=(2j+1)H_{j}^{\pm}\,,
\end{equation}
which means that the conserved charges $H^{\pm}_j$ transform with weight $2j+2$ under anisotropic scaling, in agreement with \eqref{sca}.

\section{Concluding remarks}
\label{concl}

We have performed a Hamiltonian reduction of General
Relativity in 3D with negative cosmological constant in the case of a new family of boundary conditions, labeled by a non negative integer $n$, which are related to the KdV hierarchy of integrable systems. We then obtained the action of the corresponding dual theory at the boundary, being such that the chiral copies of the reduced system evolve according to 
the potential form of the modified Korteweg--de
Vries equation \eqref{ksks}.

The asymptotic symmetries
in the bulk are then translated into Noether symmetries of the dual theory, giving rise to an infinite set of commuting conserved charges, 
that imply integrability of the system.
Remarkably, the dual action is also invariant under anisotropic Lifshitz scaling with dynamical exponent
$z=2n+1$.

It is worth pointing out that, if left and right copies were chosen according to different members of the hierarchy, the dual action turns out to be given by 

\begin{equation}
I[\varphi_+,\varphi_-]=\int d\phi dt\;\left(\frac{\ell}{32\pi G}(\dot{\varphi}_{+}\varphi'_{+}-\dot{\varphi}_{-}\varphi'_{-})-\cH_{n}^{+}[\cL^{+}]-\cH_{m}^{+}[\cL^{-}]\right)\,,
\end{equation}
where $\cL^{\pm}$ is defined in \eqref{miura}.    The anisotropic scaling symmetry would then be generically broken unless $n=m$.

It is interesting to make an interpretation of our  work in the context of the fluid/gravity correspondence
\cite{Rangamani:2009xk,Hubeny:2010wp,Bredberg:2011jq}. In that
setup, the asymptotic behavior of the Einstein equations, in a derivative expansion at the boundary, implies that the  fluid equations are recovered from the conservation of the suitably regularized Brown-York stress-energy tensor.  It is then natural to wonder about the fundamental degrees of freedom and the precise form of the theory from which the fluid is made of.

In our
context, since Einstein gravity in 3D is devoid of local propagating degrees of freedom, the identification of the fundamental degrees of freedom at the boundary can be completely performed. Indeed, the asymptotic behavior of the Einstein equations, with the boundary conditions in \cite{Perez:2016vqo}, is such that they reduce to the equations of the KdV hierarchy to all orders, i.e., without the need of performing a (hydrodynamic) derivative expansion at the boundary. Remarkably, the dynamics of the non-linear fluid, that evolves according to the KdV equations, was shown to emerge from the conservation law of left and right momentum densities, where the underlying fields are manifestly unveiled and fulfill the potential modified KdV equations.

As an ending remark,  it is worth mentioning that  different classes of  boundary conditions relating three-dimensional gravity with integrable systems have been proposed in \cite{Afshar:2016wfy,Afshar:2016kjj,Fuentealba:2017omf,Melnikov:2018fhb}. It would be interesting to explore  whether a similar construction, as the one performed here, could be carried out in those cases.

\acknowledgments We thank Oscar Fuentealba, Wout Merbis, Alfredo
Pérez, Miguel Riquelme and David Tempo, for useful discussions and comments. The work
of H.G. is supported by the Austrian Science Fund (FWF), project P
28751-N2. The work of J.M. was supported by the ERC Advanced Grant
``High-Spin-Grav'', by FNRS-Belgium (convention FRFC PDR T.1025.14
and convention IISN 4.4503.15). This research has been partially supported by FONDECYT
grants Nº 1161311, 1171162, 1181496, 1181628, and the grant CONICYT PCI/REDES 170052. The Centro de Estudios Científicos (CECs)
is funded by the Chilean Government through the Centers of Excellence
Base Financing Program of Conicyt.

\appendix

\section{Gelfand-Dikii polynomials and Hamiltonians}

\label{GDandHam}

The Gelfand-Dikii polynomials can be constructed from the recurrence
relation \eqref{recurrence}. In our conventions, the first five polynomials are
explicitly given by 
\begin{equation}
\begin{split}
R_{0}^{\pm} & =1,\label{etan}\\
R_{1}^{\pm} & =\cL^{\pm},\\
R_{2}^{\pm} & =\frac{3}{2}{\cL^{\pm}}^{2}-\frac{k}{4\pi}\d_{\phi}^{2}\cL^{\pm},\\
R_{3}^{\pm} & =\frac{5}{2}{\cL^{\pm}}^{3}-\frac{5k}{8\pi}(\d_{\phi}\cL^{\pm})^{2}-\frac{5k}{4\pi}\cL^{\pm}\d_{\phi}^{2}\cL^{\pm}+\frac{k^{2}}{16\pi^{2}}\d_{\phi}^{4}\cL^{\pm}\,,\\
R_{4}^{\pm} & =\frac{35}{8}{\cL^{\pm}}^{4}-\frac{35k}{8\pi}\cL^{\pm}(\d_{\phi}\cL^{\pm})^{2}-\frac{35k}{8\pi}{\cL^{\pm}}^{2}\d_{\phi}^{2}\cL^{\pm}+\frac{21k^{2}}{32\pi^{2}}(\d_{\phi}^{2}\cL^{\pm})^{2}\,\\
 & +\frac{7k^{2}}{8\pi^{2}}(\d_{\phi}\cL^{\pm})(\d_{\phi}^{3}\cL^{\pm})+\frac{7k^{2}}{16\pi^{2}}\cL^{\pm}\d_{\phi}^{4}\cL^{\pm}-\frac{k^{3}}{64\pi^{3}}\d_{\phi}^{6}\cL^{\pm}\,.
\end{split}
\end{equation}
Their corresponding densities \eqref{h1} then read
\begin{equation}
\begin{split}
\cH_{0}^{\pm} & =\cL^{\pm},\label{H}\\
\cH_{1}^{\pm} & =\frac{1}{2}{\cL^{\pm}}^{2},\\
\cH_{2}^{\pm} & =\frac{1}{2}{\cL^{\pm}}^{3}+\frac{k}{8\pi}(\d_{\phi}\cL^{\pm})^{2},\\
\cH_{3}^{\pm} & =\frac{5}{8}{\cL^{\pm}}^{4}+\frac{5k}{8\pi}\cL^{\pm}(\d_{\phi}\cL^{\pm})^{2}+\frac{k^{2}}{32\pi^{2}}(\d_{\phi}^{2}\cL^{\pm})^{2}\,,\\
\cH_{4}^{\pm} & =\frac{7}{8}{\cL^{\pm}}^{5}+\frac{35k}{16\pi}{\cL^{\pm}}^{2}(\d_{\phi}\cL^{\pm})^{2}+\frac{7k^{2}}{32\pi^{2}}\cL^{\pm}(\d_{\phi}^{2}\cL^{\pm})^{2}+\frac{k^{3}}{128\pi^{3}}(\d_{\phi}^{3}\cL^{\pm})^{2}\,.
\end{split}
\end{equation}
Thus, according to eq. \eqref{eomD}, with $\mu^{\pm}=R^{\pm}_n$,  the first four equations of the KdV hierarchy are given by 
\begin{equation}
\begin{split}
\pm\ell\d_{t}\cL^{\pm} & =\d_{\phi}\cL^{\pm}\,,\label{eccs}\\
\pm\ell\d_{t}\cL^{\pm} & =3\cL^{\pm}\d_{\phi}\cL^{\pm}-\frac{k}{4\pi}\d_{\phi}^{3}\cL^{\pm}\,,\\
\pm\ell\d_{t}\cL^{\pm} & =\frac{15}{2}{\cL^{\pm}}^{2}\d_{\phi}\cL^{\pm}-\frac{5k}{2\pi}\d_{\phi}\cL^{\pm}\d_{\phi}^{2}\cL^{\pm}-\frac{5k}{4\pi}\cL^{\pm}\d_{\phi}^{3}\cL^{\pm}+\frac{k^{2}}{16\pi^{2}}\d_{\phi}^{5}\cL^{\pm}\,,\\
\pm\ell\d_{t}\cL^{\pm} & =\frac{35}{2}{\cL^{\pm}}^{3}\d_{\phi}\cL^{\pm}-\frac{35k}{8\pi}(\d_{\phi}\cL^{\pm})^{3}-\frac{35k}{2\pi}\cL^{\pm}\d_{\phi}\cL^{\pm}\d_{\phi}^{2}\cL^{\pm}-\frac{35k}{8\pi}{\cL^{\pm}}^{2}\d_{\phi}^{3}\cL^{\pm} \\
 & +\frac{35k^{2}}{16\pi^{2}}\d_{\phi}^{2}\cL^{\pm}\d_{\phi}^{3}\cL^{\pm}+\frac{21k^{2}}{16\pi^{2}}\d_{\phi}\cL^{\pm}\d_{\phi}^{4}\cL^{\pm}+\frac{7k^{2}}{16\pi^{2}}\cL^{\pm}\d_{\phi}^{5}\cL^{\pm}-\frac{k^{3}}{64\pi^{3}}\d_{\phi}^{7}\cL^{\pm}\,.
\end{split}
\end{equation}

\section{pmKdV equations}
\label{modi}

The first four equations of the potential modified KdV hierarchy read
\begin{equation}
\begin{split}
\pm\ell\dot{\varphi}_{\pm} & =\varphi_{\pm}'\,,\label{eqpmKdV}\\
\pm\ell\dot{\varphi}^{\pm} & =\frac{k}{4\pi}\left(\frac{1}{2}{\varphi_{\pm}'}^{3}-\varphi_{\pm}'''\right)\,,\\
\pm\ell\dot{\varphi}^{\pm} & =\left(\frac{k}{4\pi}\right)^{2}\left(\frac{3}{8}{\varphi_{\pm}'}^{5}-\frac{5}{2}\varphi_{\pm}'{\varphi_{\pm}''}^{2}-\frac{5}{2}{\varphi_{\pm}'}^{2}\varphi_{\pm}'''+\varphi_{\pm}^{(5)}\right)\,,\\
\pm\ell\dot{\varphi}^{\pm} & =\left(\frac{k}{4\pi}\right)^{3}\left(\frac{5}{16}{\varphi_{\pm}'}^{7}-\frac{35}{4}{\varphi_{\pm}'}^{3}{\varphi_{\pm}''}^{2}-\frac{35}{8}{\varphi_{\pm}'}^{4}{\varphi_{\pm}'''}+\frac{35}{2}{\varphi_{\pm}''}^{2}{\varphi_{\pm}'''}+\frac{21}{2}{\varphi_{\pm}'}{\varphi_{\pm}'''}^{2}\right. \\
 & \left.\quad\quad\quad\quad\quad\quad+14{\varphi_{\pm}'}{\varphi_{\pm}''}{\varphi_{\pm}''''}+\frac{7}{2}{\varphi_{\pm}'}^{2}{\varphi_{\pm}^{(5)}}-{\varphi_{\pm}^{(7)}}\right)\,.
\end{split}
\end{equation}

\bibliographystyle{fullsort}

\end{document}